\documentclass[12p]{article}
\usepackage{latexsym}
\usepackage{color}
\usepackage{amsmath}
\usepackage{epsfig}
\usepackage{hyperref}
 \usepackage{dcolumn}
\newcommand{\ortala}[1]{\begin{center}#1\end{center}}

\newcommand{\sandd}[1]{\left\langle #1\right\rangle}

\newcommand{\integ}[3]{{{\underset{#1 }{\overset{#2}{\displaystyle\int}}}#3}}

\newcommand{\re}[1]{(\ref{#1})}

\newcommand{\eq}[2]{\begin{equation}\label{#1}  #2\end{equation}}

\newcommand{\paran}[1]{\left(#1\right)}

\newcommand{\sch}[1]{Schrodinger}

\usepackage{amssymb}
\usepackage{latexsym}
\begin{document}

\ortala{\large\textbf{Testing the Goodwin Growth Cycles with Econophysics Approach 
in 2002-2019 Period in Turkey}}

\ortala{\textbf{Kerim Eser Af\c{s}ar*, Mehmet \"Ozyi\~git*, Yusuf Y\"uksel**, \"Umit Ak\i nc\i** \footnote{\textbf{umit.akinci@deu.edu.tr}}}}

\ortala{\textit{*Department of Economics, Dokuz Eyl\"ul University,
TR-35160 Izmir, Turkey\\
**Department of Physics, Dokuz Eyl\"ul University,
TR-35160 Izmir, Turkey}}

\section{Abstract}

The Turkish economy between 2002-2019 period has been investigated within the econophysical approach. From the individual income data obtained from the Household Budget Survey, the Gompertz-Pareto distribution for each year and Goodwin cycle for the mentioned period have been obtained. For this period, in which thirteen elections were held under the single-party rule, it has been observed that the income distribution fits well with the Gompertz-Pareto distribution which shows the two-class structure of the Turkish economy. The variation of the threshold value $x_t$  (which separates these two classes) as well as Pareto coefficient have been obtained. Besides, Goodwin cycle has been observed within this period, centered at $(u,v)\cong (66.30,83.40)$ and a period of  $T=18.30$ years. It has been concluded that these observations are consistent with the economic and social events experienced in the mentioned period.

\section{Introduction}\label{Introduction}

It can be considered as an obvious reality in terms of social sciences that income and wealth are not equally and homogeneously distributed in a society. The inequality of income and wealth distribution is handled with some factual or quantitative classifications. However, some qualitative questions regarding the income and wealth distribution remain as uncovered, such as how income and wealth are distributed, what is the functional form of the distribution, whether this distribution has a universal form (character) or it displays a country-specific character, whether it depends on a particular time and historicity. In this regard, the strong regularity shown by the income data (top $1-10\%$ and remaining $99-90\%$) has opened up space for studies revealing the characteristics of the statistical distribution, taking into account the extreme values in the distribution of the data \cite{ref1,ref2}. The first studies on income distribution in economic theory are based on the studies of Pareto \cite{ref3} in which, he tried to explain the income distribution with a universal law that was valid for all times and all countries. In the studies carried out afterwards, it was stated that the base law fluctuates in a certain interval depending on the time and the country. Then, with the emergence of the idea that the income distribution in a particular country cannot be explained by a single type of distribution, the literature of econophysics has been enriched by the efforts of economists and/or physicists to determine the exact shape of the distribution based on raw data and to create theoretical models that can reproduce these distributions.

At this point, econophysics has emerged as a new interdisciplinary research area that uses the laws explained, and the theories and methods developed by physicists for the solution of the economic problems which has uncertainty, stochastic processes and nonlinear dynamics. For the purpose of concrete analysis of economic reality, it tries to explain economic causes and effects with physical causes and effects. The fundamental tools of econophysics consist mostly of probabilistic and statistical methods taken from statistical physics. Econophysics is an interdisciplinary area that applies the ideas, models, concepts and methods of statistical physics to quantitative economics \cite{ref4,ref5,ref6,ref7,ref8,ref9,ref10,ref11,ref12,ref13}.

Although, the use of income as a class indicator in realizing social reproduction of individuals under capitalist relations is a controversial issue, income is an acceptable indicator in terms of quantitative analysis. Therefore, the community which has different distribution patterns should be analyzed by dividing it into classes, sub-layers or groups in some way, as well as these different distribution numbers. Depending on the historical and social dynamics in the literature, it can be seen in the context of various country and period cases where the distribution below and above a certain threshold value occurs functionally in different ways. Grasping of the economic inequality, power law in particular has emerged as a critical concept because of the connections with consumption, business cycles and other macroeconomic phenomena. In the literature of econophysics, there are a large number of studies that have analyzed income inequality on the basis of country, period and variable. Some of these works can be seen in Table \ref{tbl_1}.

\begin{table}\caption{Econophysics and Literature of Income Distribution.
Abbreviations in  column source: S.H.-Size of Houses; I-Income; I.T.-Income Tax; Inhe. T.-Inheritance Tax; I.R.-Inland Revenue; W.-Wealth.
Abbreviations in column distribution : L.N.-log-normal; Exp.-Exponential (the same with Boltzman-Gibbs in Physics).
}
\scriptsize
 \begin{tabular}{|c|c|c|c|c|}\label{tbl_1}
 \textbf{Country}& \textbf{Source}& \textbf{Distribution}& \textbf{Pareto Exponent}&\textbf{Ref.}\\
 \hline
 Egypt & S.H. (14th B.C.)&Pareto&$\alpha=1.59 \pm 0.19 $&\cite{ref14}\\
  \hline
 Japan & I. (1997) and I.T. (1997-8)& Pareto&$\alpha=-1.98$ and $\alpha=-2.05$&\cite{ref15}\\
 \hline
 Japan &I. (1987-2000)&Pareto&$\alpha=1.8-2.1$&\cite{ref16}\\
 \hline
  USA &I. (1935-36) & L.N./Pareto& $\alpha=1.63$&\cite{ref17}\\
 \hline
UK &Inhe. T. (2001) &Pareto& $\alpha=1.78$&\cite{ref18}\\
 \hline
UK & IR. (1996) & Exp./Pareto& $\alpha=1.9$&\cite{ref19,ref20}\\
USA&  I. (1998)& & $\alpha=1.7\pm 0.1$&\\
 \hline

UK &I. (1992-2002) &L.N. / Pareto& $\alpha=[3.34-2.68]$&\cite{ref1}\\
 \hline
Germany, &I. (1990-2002) &L.N. (Gibrat index,$\beta$) & $\alpha=[2.42-3.96]$&\cite{ref21}\\
& &/ Pareto& $ \beta=[1.63-2.14]$&\\

UK, &I. (1991-2001) & & $\alpha=[3.47-5.76]$&\\

 & & & $\beta=[2.18-2.73]$&\\

USA &I. (1980-2001) & & $\alpha=[1.1-3.34]$&\\

&& & $ \beta=[1-1.65]$&\\

 \hline
Australia &I. (1993-1997) &L.N./ Pareto& $\alpha=[2.2-2.6]$&\cite{ref22}\\
 \hline
Italy&I. (1977-2002) &Two parameter L.N.& $\alpha\cong 2.90, \beta\cong  2.30$&\cite{ref23}\\

& & / Pareto& &\\

 \hline
Columbia&I. (2010-Q2, Q3,&Exp. / Pareto& $\alpha\cong  2.5$&\cite{ref24}\\
& 2011-Q2, 2012-Q4)&& &\\

 \hline
India&W. (2002-04)&Pareto& $\alpha\cong  0.81-0.92$&\cite{ref25}\\

&I. (1997)&Pareto& $\alpha\cong  1.5$&\\
 \hline
Brazil&I. (1981-2009)&Gompertz - Pareto& $\alpha\cong  2.75, \beta \cong  0.35$&\cite{ref26,ref61,ref62}\\

 \hline

 \end{tabular}

\end{table}

In the econophysics literature where income distribution is examined, contrary to general acceptance, the log-normal distribution is not sufficient to explain the whole process. While income distribution under a certain threshold value is adjusted to distributions such as log-normal, exponential or Gompertz, the Pareto distribution is used to explain the extreme values. Thus, it is stated that income distribution exhibits two different regimes \cite{ref21,ref22}. Therefore, a socially two-regime structure in the analysis of income distribution and economic inequalities is frequently used. This two-regime structure reveals that a two-class social structure must be considered in examination of economic relations: one part of the distribution that refers to the low-income earners under the threshold value, those who earn labor income and the other part refers to those who earn capital incomes. Moreover, the two-regime structure of income distribution underlies the necessity of determining the different dynamics of the low and high levels of distribution in the context of economic inequalities. On the other hand, evolution of the distributions of these two different regimes by the years, gives important clues about the investigated economical system as a dynamical system. 

Another dynamic character of the system can be explained by Goodwin cycles. In the Goodwin model \cite{ref28}, the dynamics of Lotka-Volterra prey-predator model are defined with two new ($u$ and $v$) variables in economic context. Where $u$ stands for the share of workers in total production, which is an indirect way of explaining the profit margin of capitalists. The second term denoted as $v$ defines the employment rate, which is an indirect way of explaining the share of unemployed workers or the industrial reserve army in Marx's terminology. This dynamic cyclical process of economic expansion can be explained by two variables as follows: with the realization of capital accumulation or growth and, the employment rate $v$ increases. The share of workers begins rising with a certain delay. This means a reduction or a squeeze in the profit margin. When the employment rate reaches its maximum then the profit margins reduce to their lowest level. The decrease in capital accumulation firstly causes a decrease in the employment rate. At this point, the share of labor in total output/yield already tends to decrease. The model reveals the relationships between economic variables as a closed trajectory in the $u - v$ phase space \cite{ref26}.
After Goodwin \cite{ref28} published his basic model, many studies have been carried out that test the validity of the model\cite{ref27}, trying to develop the model by extending its basic assumptions or adding theoretical contributions \cite{ref28,ref29,ref30,ref31,ref32,ref33,ref34,ref35,ref48,ref50}, and examining the empirical results of the model \cite{ref26,ref36,ref37,ref38,ref39,ref40,ref51,ref57,ref58}. Studies conducted in the context of Turkey also include testing the periodic validation of the model with different methods \cite{ref41,ref42,ref43,ref44,ref45}.

Although the above mentioned theoretically well-established Gompertz-Pareto Distribution (GPD) and Goodwin cycles are consistent and well understood conceptually, we believe that, they still need to be verified by applying them on more examples. The first motivation of this study is to demonstrate the applicability of these tools once again, but for the Turkish economy. On the other hand, the Turkish  economy since 2002 (which is the starting year of the investigation presented in this study) needs more elaboration. We believe that, by its uninterrupted single-party government, two major economic crises, one social rebellion (Gezi Park), a coup attempt and many elections, Turkey is an interesting example \c sn terms of the economy as a dynamical system. The effect of some external stimulants on the evolution of an economical system (as a dynamical system) is still not sufficiently understood. Therefore, it is valuable to determine the effect of all these events in the relevant period on the Turkish economy. This is the second motivation of our study.

From these motivations, we investigate the Turkish economy between the 2002-2019 period within the econophysical approach, by using the Household Budget Survey (HBS) which contains income data of individuals between the period 2002-2019.  To the best of our knowledge, this study is the first study attempting to calculate the Gompertz-Pareto Distribution (GPD) based on the result of the fitting process and the Pareto exponent and the Gini coefficient based on the two-class character of the income distribution for the Turkish economy. Besides, the Goodwin growth cycle model which considers the dynamics in a capitalist society as internal relations was used to examine the inequality dynamics of the Turkish economy for the period of 2002-2019. Although there are some models which examine Goodwin cycles and its different extensions, we prefer to follow the literature that deals with Goodwin cycles on the basis of econophysics. We note that, Goodwin model was reconstructed on the basis of Ref. \cite{ref26}.

The paper is organized as follows: the basic assumptions and mathematical form of the Goodwin model, as well as GPD are explained in section \ref{sec_formulation}. Calculation results regarding the distributional structure of the data, and the Goodwin cycle of the Turkish economy for the period 2002-2019 as well as discussions on them are presented in section \ref{sec_results}. Finally, section \ref{sec_conclusion} presents of our conclusions.

\section{Model and Formulation}\label{sec_formulation}

We analyse the income distribution in Turkey for income data between years 2002-2019 which is obtained from the Household Budget Survey. As a macroeconomic model we use Goodwin growth-cycle model. Besides, we construct the GPD for each year and by this way  we obtain Goodwin cycle for the mentioned period.

\subsection{The Goodwin Growth-Cycle Macroeconomic Dynamics}

The capitalist accumulation process takes place depending on the organic composition of capital and the course of the organic composition in the process of accumulation \cite{ref46}. This process causes a compulsory and conflicting accumulation relationship between the fixed capital or value of means of production and the variable capital or the value of labor force. The phases of expansion and contraction in the capital accumulation process reveal the cyclical characteristics of the contradictions between labor and capital. During the expansion phase of capital accumulation, the increase in production and the rise in profits cause to decrease in unemployment and the shortage of labour supply relatively to capital accumulation leads to better and/or higher-paid job opportunities for the labor. During the contraction periods after the expansion, the reduction in production and the decrease in profits create an increase in unemployment, an expansion of the industrial reserve army, thereby suppressing wages and causing a decline in workers' bargaining power against capital. As the decline in wages causes increase in  profit opportunities and create new investment opportunities, it will initiate a new expansion phase. Thus, the cyclical structure of accumulation will emerge \cite{ref26,ref47,ref48}. In this context, the main difference of the form of capital accumulation process propounded by Marx is that investments, unemployment and technological change are internalized into the growth process as economic factors that both affect and are affected by capital accumulation.

The dynamic economic relations propounded by Marx were formulated by Goodwin, depending on their theoretical roots and in a simple manner \cite{ref49}. Goodwin model is a theory that explains the internal cycles generated by distributive conflicts between classes in which economic variables interact with each other cyclically and internally in economic growth cycles in which cycles are not the result of external shocks due to chance or condition but the dynamic interaction of deterministic variables of the model. The model derives dynamic relations mathematically by applying a specific functional form to behavioral characteristics of capitalists and workers and the production function. The contradiction between wages and profits is explained by nonlinear differential equations, with a dynamic accumulation model in the space of wage share-employment rate. The essential dynamics of capitalist society have been established with the adaptation of the classic Lotka-Volterra prey-predator model to economic relations in which capital and labor are conflicted but symbiotic, as a single-sector and two-class model of share struggle in national income \cite{ref48,ref50,ref51,ref52}.

\subsection{Lotka-Volterra prey-predator model}
Lotka-Volterra (LV) systems, independently modeled by Lotka \cite{ref53} and Volterra \cite{ref54}, describe population dynamics in ecological systems \cite{ref55,ref56}. In the LV model, it is possible to examine the coexistence relations of living beings in a particular flora. The interaction between preys and predators is modeled in this framework. Preys (which has population $x$) are assumed to have unlimited feeding possibilities and that their population increases at the rate of $a_2$.
 On the other hand, under the presence of predators, the prey population decreases at the rate of $-b_2y$, where $y$ is the population of the predators. Similarly, predators  have a natural death rate of $-a_1$. Under the presence of preys, the predator population also increases at the rate of $b_1 x$. The process is based on explaining population dynamics based on the interaction between these two species. Under the assumption that there are no predators in the system and there is a regular food source for preys, the prey population increases at an exponential rate with the exponent $a_2$. On the other hand, by assuming that there is no prey in the ecological system, the predator population decreases exponentially with the exponent $a_1$. In the ecological system, when preys and predators come together, the growth dynamics of the populations in the system is defined by LV differential equations which are given by,

\eq{denk1}{
\begin{array}{lcl}
\frac{dx}{dt}&=& x(a_2-b_2 y)\\ 
\frac{dy}{dt}&=& y(-a_1+b_1 x).
\end{array}
}

By adapting the prey-predator dynamics in the LV model, Goodwin uses variable $u$ variable to refer to the share of workers in total production as an indirect form of describing the profit margins of capitalists. Goodwin also uses the variable $v$ to express the rate of employment as an indirect form of explaining the industrial reserve army in Marx's terminology so the share of the unemployed \cite{ref25,ref57,ref58}. The model expresses the continuity of a permanent industrial reserve army in the interaction of wages, profits and unemployment. Besides, it is expressed as a growth cycle model in which profits compressed by the laborer wages as a result of different contributions are converted into an investment, and thus the speed of capital accumulation is determined \cite{ref57,ref58}.

\subsection{Mathematical Framework of Goodwin Growth Cycle Model}

Goodwin \cite{ref49,ref59}, explains dynamic relations between wage share ($u$) and employment rate ($v$) with the differential equations: 

\eq{denk2}{
\begin{array}{lcl}
\dot {v} /v &=& \left[\left(1/\sigma- \left(\alpha+\beta \right) \right) - \left(1/\sigma \right) u \right]\\ 
\dot {u} /u  &=& [ - (\alpha+\gamma) +\rho v ].
\end{array}
}
Here, $1/\sigma$ the productivity of capital, $\alpha$ is the labor productivity growth rate, $\beta$ in the population growth rate, $\rho$ represents the sensitivity of the change in real wages to employment
and, $\gamma$ is the constant of Phillips curve, which contains information about the rates of unemployment and corresponding rates of rises in wages.

On the other hand, the model defines the change in nominal wages as a function of the employment rate and adds a Phillips curve with real wages ($w$) to the model as
\begin{equation}\label{denk3}
\dot{w}/w=-\gamma+\rho v.
\end{equation}

The above equation implies that the higher the employment rate, the faster the real wages will grow. Since $u$ as the share of wages is defined by the ratio of real wages to labour productivity, the above equation constitutes the third basic equation of the model as
\begin{equation}\label{denk4}
\dot {u} /u =  \dot {w} /w - \alpha = - (\alpha+\gamma)+\rho v.
\end{equation}
Here $\rho$ shows the slope of the Phillips curve and the effect of increase in employment on real wages, because in Goodwin's model, employment growth also increases the wage bargaining power of the working class, leading to a jump in real wages. A positive shock to the bargaining coefficient will lead to the layoffs of workers and thus a decrease in the equilibrium employment rate. Similarly, a positive shock to productivity must be compensated by higher profit rates and lower wage share \cite{ref56}.

It is trivial that, the equations that defines Goodwin model in Eq. \re{denk2} and equations of LV given in Eq. \re{denk1} are in the  same form. It can be seen by comparing the aforementioned system of equations provided that the relations
\eq{denk5}{
\begin{array}{lcl}
a_1  = \alpha+\gamma&,& \quad b_1 =\rho \\
a_2 =1/\sigma - (\alpha+\beta) &,& \quad b_2  = 1/\sigma\\
\end{array}
}
can be established, then Goodwin model can be treated as LV model. Indeed, these conceptual relations exist and  thus, the two sets of equations expressed by $(u,v)$ in Eq. \re{denk2} and $(x,y)$ in Eq. \re{denk1} formally show the Goodwin growth cycle model in accordance with the LV predator model. In this treatment, while  employment rate $v$ in Goodwin model corresponds to the preys, labor share $u$ corresponds to the predators. Let us give some important points about the relations between the LV model and Goodwin model. These relations  are important in order to  decide whether the model could explain the economic reality or not.

According to Eq. \re{denk5}, $a_2$ denotes the difference between the sum of the labor productivity growth rate and the population growth rate and the productivity of capital. This coefficient, which is the growth rate of the prey population in the context of the LV equations, refers to the growth in the employment rate in the context of the Goodwin growth cycle, when the workers do not demand any product from the total product.  $b_2$
denotes a decrease in the prey population. Therefore, under the current conditions, it will show the decrease in employment rate because of the increase in the productivity of the capital. While $a_1$ is the natural death rate of predators in the context of LV,  it is economically the sum of the rate of increase in labour productivity and the constant of the Phillips curve. $b_1$ means the sensitivity of the change in real wages to employment, which is responsible for the increasing behavior in the population of the predators.

Under certain initial conditions, and for the values of the parameters in Eq. \re{denk5}, $u-v$ relation reveals repeated cycles. Employment grows at the same rate as the working population in the long run, as the employment rate fluctuates around an average level. Therefore, output grows at a natural rate that is the sum of population growth and productivity growth in the long run \cite{ref60}. Also, since $u$ denotes the share of labor, it implies that the real wage fluctuates around a rising trend at the same rate as labour productivity. A cycle (or repeated cycles) constructed from Eq. \re{denk2} 
has a center \cite{ref36} 
\eq{denk6}{
v_c=\frac{a_1}{b_1}, \quad u_c=\frac{a_2}{b_2},
}
and a period
\eq{denk7}{
T=\frac{2\pi} {\sqrt{a_1a_2}}.
}

\subsection{The Gompertz-Pareto Income Distribution}

We assume that the Turkish economy consists of two main classes which are, workers and capitalists, as valid for many modern capitalist societies. Workers share the lower part of the distribution, and capitalists share the remaining higher part of the income. The workers' share can be given by a Gompertz curve, while the capitalists' share are represented by Pareto distribution. Thus, the cumulative income distributions (which define the probability that an individual receives an income less than or equal to $x$) of our model are as follows:
\begin{equation}\label{denk8}
\mathcal{F}(x)=\left\{
\begin{array}{lcl}
 \mathcal{G}(x)&,&0\leq x<x_t\\
 \mathcal{P}(x)&,&x_t\leq x \leq \infty\\
\end{array}
\right.\end{equation}
where the function in the Gompertz region is given by

\begin{equation}\label{denk9}
\mathcal{G}(x)=1-\exp{\left[-\eta\paran{ e^{bx}-1}\right]}
\end{equation}
while for the Pareto region we have
\begin{equation}\label{denk10}
\mathcal{P}(x)=1-\paran{x_t}^\alpha \exp{\left[-\eta\paran{ e^{bx_t}-1}\right]}x^{-\alpha}.
\end{equation}
Here $\alpha$ is the Pareto index and $x_t$ is the income threshold value  of the Pareto region, in other words it is the value of income that separates the working class and capitalist class. The Pareto form of the distribution is nothing but the $\beta x^{-\alpha}$  which satisfies continuity at $x=x_t$ with the Gompertz distribution. Note that the condition $\mathcal{G}(0)=0$ is automatically satisfied.

The probability densities of income distributions are defined via:
\begin{equation}\label{denk11}
f(x)=\left\{
\begin{array}{lcl}
 g(x)&,&0\leq x<x_t\\
 p(x)&,&x_t\leq x \leq \infty\\
\end{array}
\right.
\end{equation}
where

\begin{equation}\label{denk12}
g(x)=\eta b e^{bx}\exp{\left[-\eta\paran{ e^{bx}-1}\right]}
\end{equation}
and

\begin{equation}\label{denk13}
p(x)=\alpha \paran{x_t}^\alpha \exp{\left[-\eta\paran{ e^{bx_t}-1}\right]}x^{-\alpha-1}
\end{equation}
are distributions for Gompertz and Pareto regions, respectively. This $f(x)$ defines the probability that an individual has income between the $x$ and $x+dx$. Equivalently $f(x)$ is a fraction of individuals with income
between $x$ and $x + dx$.

Normalization of these probability densities requires,
\begin{equation}\label{denk14}
\integ{0}{\infty}{}f(x)dx=\integ{0}{x_t}{}g(x)dx+\integ{x_t}{\infty}{}p(x)dx=1
\end{equation}
holds. 
Mean value of the income for the whole individuals is defined by 
\begin{equation}\label{denk15}
\sandd{x}=\integ{0}{\infty}{}x f(x)dx=\left[I\paran{x_t}+
\frac{\alpha x_t}{\alpha-1}\exp{\left[-\eta\paran{ e^{bx_t}-1}\right]}\right]\end{equation}
where
\begin{equation}\label{denk16}
I\paran{x}=\integ{0}{x}{}y \eta b 
 \exp{\left[-\eta\paran{ e^{by}-1}+by\right]}  dy
\end{equation}

In order to determine the degree of inequality of the income distribution we use Lorenz curve.  It is a two-dimensional  plot with vertical axis is the cumulative share of income $\mathcal{F}_1(x)$ and the horizontal axis is $\mathcal{F}(x)$. $\mathcal{F}_1(x)$ can be determined by the definition which is
\eq{denk17}{
\mathcal{F}_1(x)=\frac{1}{\sandd{x}}\integ{0}{x}{}yf(y)dy,
} for the two regions of income, it is given by
\eq{denk18}{
\mathcal{F}_1(x)=\frac{I(x)}{\sandd{x}}, \quad 0<x<x_t
}  and 
\eq{denk19}{
\mathcal{F}_1(x)=1+\frac{\alpha\paran{x_t}^\alpha \exp{\left[-\eta\paran{ e^{bx_t}-1}\right]}}{\paran{1-\alpha}\sandd{x}}x^{1-\alpha}, \quad x_t\leq x<\infty
}
The Gini coefficient is the measure of the inequality of the income distribution calculated as the area between the perfect equality curve and Lorenz curve. 

\eq{denk20}{
Gini=1-2\integ{0}{\infty}{}\mathcal{F}_1(x)f(x)dx.
} It can be calculated by using \re{denk11} and \re{denk17} in \re{denk20}, and is given by
\eq{denk21}{
\begin{array}{lcl}
 Gini&=&1-2\left\{\frac{\eta b}{\sandd{x}} \integ{0}{x_t}{}I(x)e^{bx}\exp{\left[-\eta\paran{ e^{bx}-1}\right]dx}+\exp{\left[-\eta\paran{ e^{bx_t}-1}\right]}\right.\\
&+&\left. \frac{\alpha^2x_t}{\sandd{x}(1-\alpha)(2\alpha-1)}\exp{\left[-2\eta\paran{ e^{bx_t}-1}\right]}
\right\}
\end{array}
}

The parameter $x_t$ and the other parameters appearing in the GPD functions given in Eqs. \re{denk12} and \re{denk13} can be determined by using nonlinear fitting procedures. 

\section{Results and Discussion} \label{sec_results}

All calculations were performed using normalized income data, which are obtained by dividing all income data in a year by the average income for that year. In other words, average value of income data in one year is taken as unity.  

Two types of calculations were performed in this work. One by fitting, and the other from the raw data. Note that, the aim of fitting is to determine the suitability of the GPD for income data of Turkey and to evaluate the value of  $x_t$ which separates two regimes of income. Once $x_t$ is determined for years, we can obtain cycles and other properties  from raw data. The results  obtained by using raw data are denoted in square brackets. For instance, while $[Gini]$ refers to calculated value of Gini coefficient from raw data, $Gini$ denotes the same coefficient calculated by the values of parameters obtained by fitting procedure (i.e. from Eq. \re{denk21}).

Note that, since all distribution functions given above produce values in interval $[0,1]$, then in order to get percentage values, we need to enlarge them by $100$.

\subsection{GPD of Income Data for Turkey}

By analyzing the income data for Turkey, we conclude that fitting data to the exponential distribution is inadequate. This fact has also been demonstrated in Refs. \cite{ref26,ref61,ref62} for the Brazilian economy. 
We start by determining the suitable form of the fitting function for a Gompertz regime, by comparing a different form of the Gompertz distribution by the form used in \cite{ref26}.

\begin{figure}[h]
\center
\epsfig{file=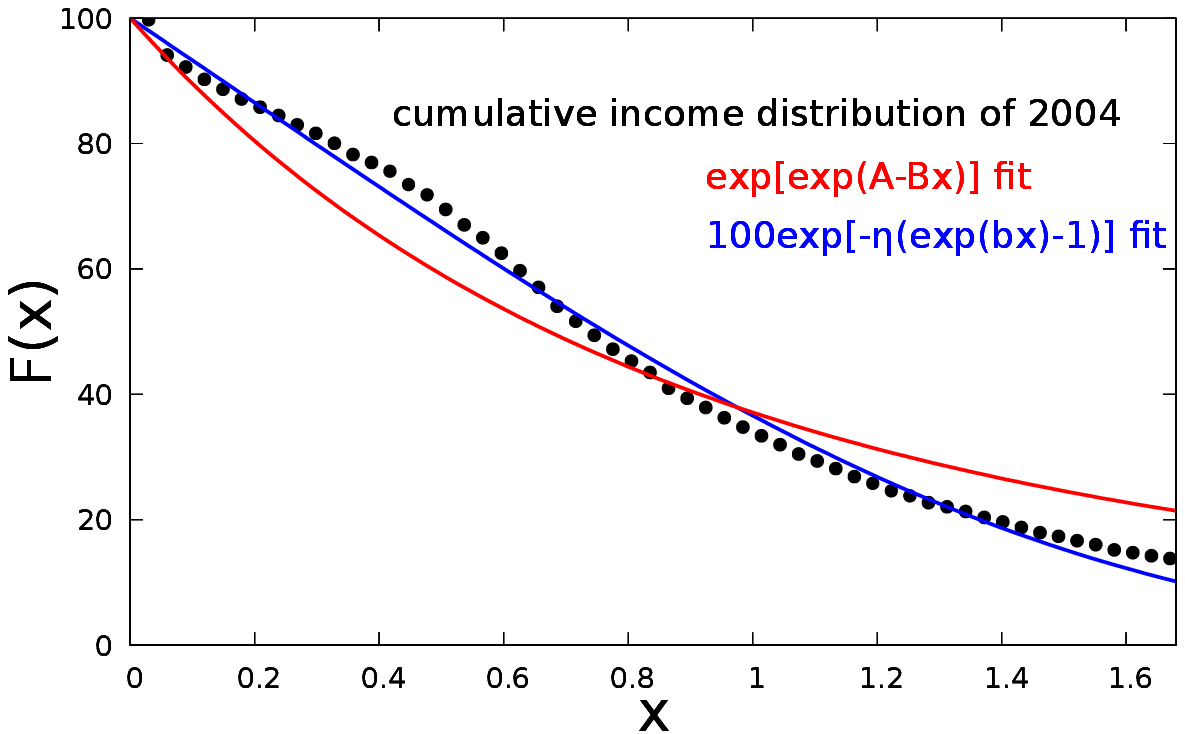, width=6cm}
\epsfig{file=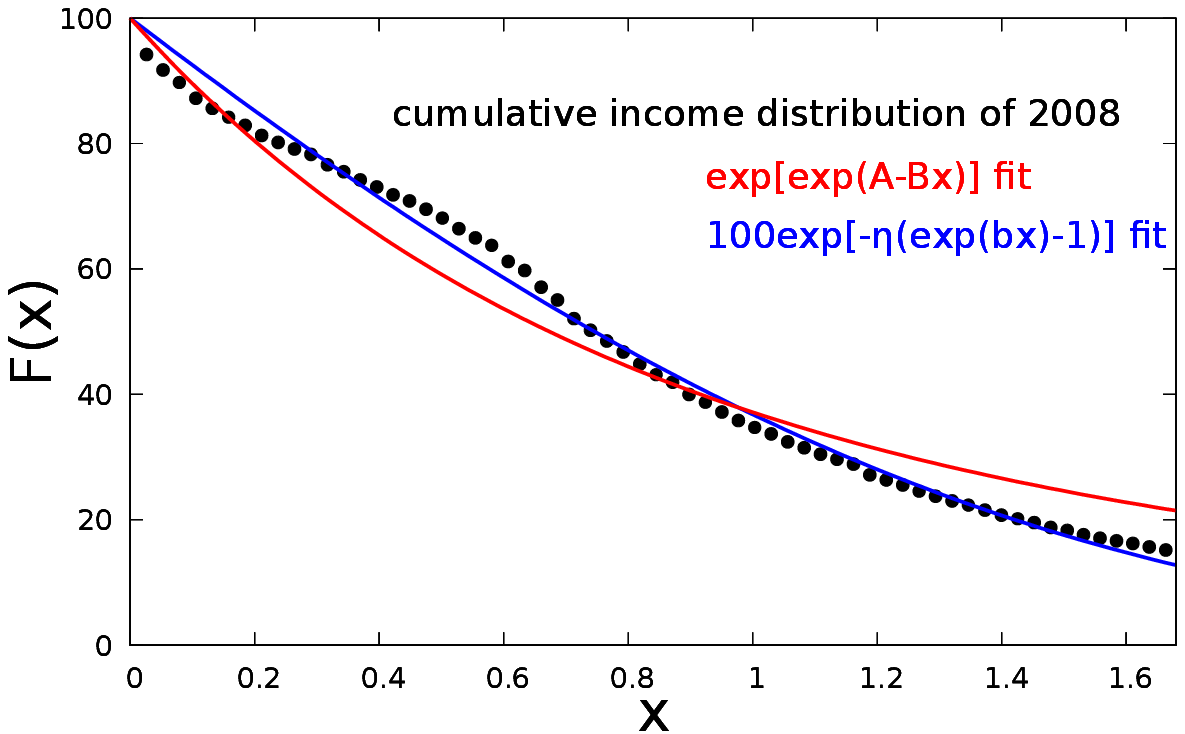, width=6cm}
\epsfig{file=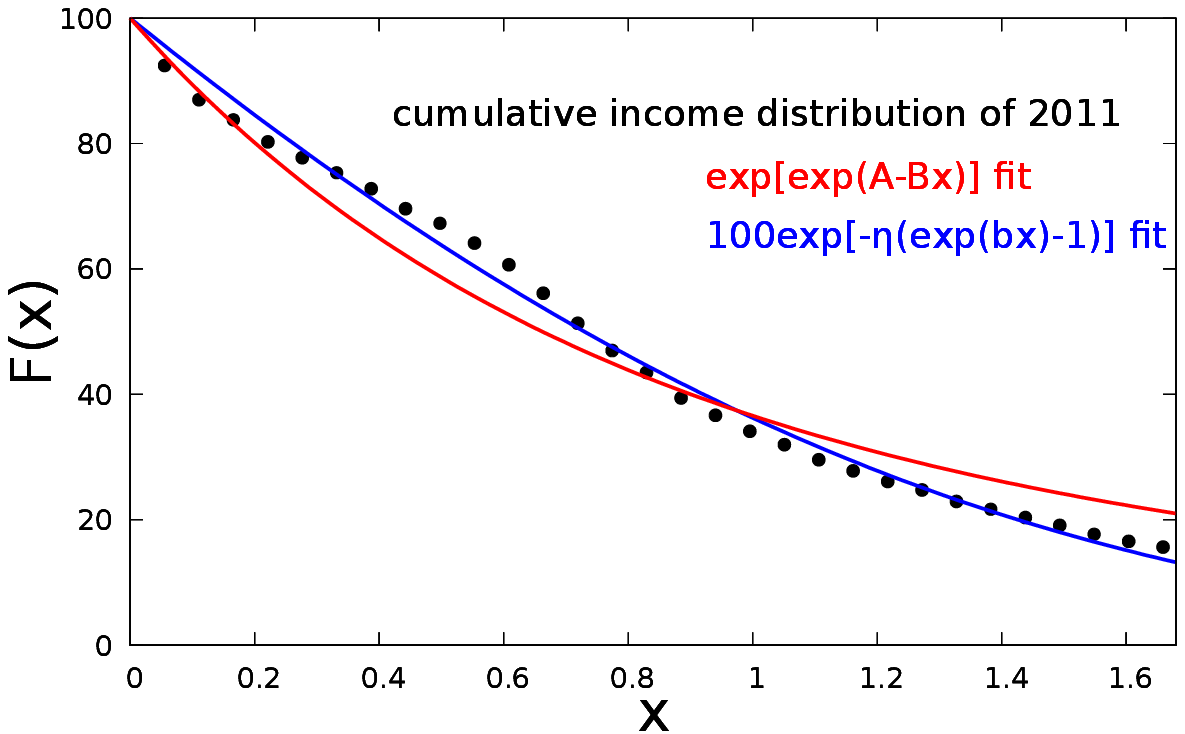, width=6cm}
\epsfig{file=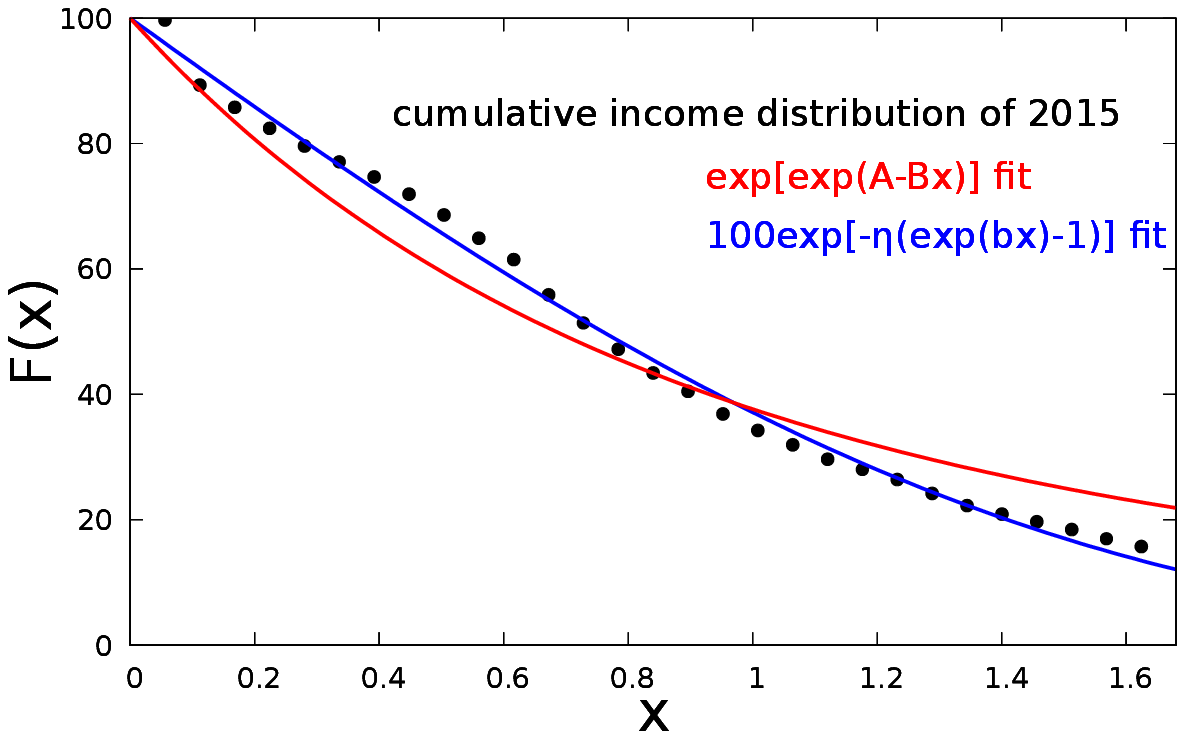, width=6cm}
\caption{The fit of Gompertz curve to Turkey's workers' individual cumulative normalized income distribution data for selected years of 2004, 2008, 2011, 2015. }\label{sek1}
\end{figure}

As seen in Fig. \ref{sek1}, the form of Eq. (9) gives more reasonable approximation than the form used in \cite{ref26}, for income distribution of the workers of some selected years. This conclusion  is valid for all years for our data.  Because of this fact,  we prefer to use a different form of the Gompertz distribution function than the form used in Ref. \cite{ref26}. In order to obtain the Goodwin cycles in the income and employment data of Turkey, we need to determine percentage share of the Gompertzian part of an income distribution $u$ and the proportion of labor force employed $v$. Since we don't have direct data for $v$, we should make some estimation for it. In this regard, we follow the methodology presented in \cite{ref26}.

After determining the specific form of the Gompertz distribution which is employed for analyzing Turkey case, the threshold value $(x_t)$ dividing income data in two parts, as well as values of the other parameters were calculated. Depending on $x_t$, we calculate the Gini coefficients from the raw data  and fitting procedure by using Eq. \re{denk21}. These values can be seen in Table \ref{tbl_1}. As a measure of the success of fitting procedure, we can mention that mean value of the income given by Eq. \re{denk15}  deviates from $1.000$ in order of $10^{-3}-10^{-4}$.  
 
\begin{table}\caption{Values of the parameters and Gini coefficients.  $x_t$ values in the table were obtained from HBS individual income according to the procedures explained above. As a result of the fit procedure performed using these values, the other parameter values in the table were obtained.
}
 \begin{tabular}{c|c|c|c|c|c|c}\label{tbl_2}
Year&$x_t$&$\eta$&$b$&$\alpha$&[Gini]&Gini\\
\hline
2002	&2.135	&2.491	&0.358	&1.598	 &0.505	 &0.533\\
2003	&1.446	&0.604	&1.006	&1.809	 &0.462	 &0.485\\
2004	&1.432	&0.691	&0.909	&1.938	 &0.469	 &0.481\\
2005	&1.722	&1.169	&0.610	&2.075	 &0.473	 &0.479\\
2006	&1.754	&1.381	&0.537&2.103	 &0.483	 &0.485\\
2007	&2.048	&1.939	&0.406	&2.172	 &0.490	 &0.484\\
2008	&1.796	&1.561	&0.493	&2.049	 &0.490	 &0.492\\
2009	&2.072	&2.921	&0.299	&2.112	 &0.509	 &0.502\\
2010	&2.041	&2.372	&0.350	&2.183	 &0.497  &0.492\\
2011	&1.936	&2.136	& 0.386	&2.085	 &0.499	 &0.497\\
2012	&1.936	&1.738	&0.452	&2.122	 &0.487	 &0.485\\
2013	&2.109	&2.146  &0.380	&2.110	 &0.491	 &0.488\\
2014	&2.062	&1.955	&0.408	&2.112	 &0.485	 &0.486\\
2015	&1.849	&1.369	&0.541	&2.131	 &0.477	 &0.476\\
2016	&1.921	&1.185	&0.586	&2.188	 &0.459	 &0.462\\
2017	&1.913	&1.001	&0.661	&2.134	 &0.456	 &0.455\\
2018	&1.705	&0.770	&0.813	&1.998	 &0.456	 &0.460\\
2019	&1.787	&0.919	&0.703	&2.256	 &0.449	 &0.451\\
    
\end{tabular}
\end{table}

As seen in Table \ref{tbl_1}, the values of the parameters related to the Gompertzian part of the income distribution ($\eta$ and $b$) has large  fluctuations in comparison with the exponent of Pareto regime ($\alpha$). 
In the vicinity of the Gompertz regime, for income levels below $x_d$, as the income of individuals increases, the probability of gaining increasing income also increases. The probability from the threshold value ($x_d$) to the minimum wage, grows decreasingly . Individuals are twice less likely to be in a certain income range than they are in the previous income range at the transition point from the minimum wage to the Pareto regime ($x_t$). Since there is no regularity mentioned in the HBS data, the parameters are not fixed in the Gompertz regime. As the rate of change of transition probabilities $b$ changes, the number of people in a certain income range ($\eta$) also changes in the opposite direction. However, the multiplication of the two coefficients in the double exponential form provides the functional form of the distribution.

Pareto coefficient $\alpha=2$ indicates that when the income level of the individuals doubles, then the number of individuals in that group decreases four times. Since the Pareto coefficient for 2002 was calculated from the 2001 data, the Pareto coefficient is increased from  $1.5$ phase to $2.0$ phase between 2002 and 2019, reaching its historically highest level $2.256$ phase in 2019. The Gini coefficient settles in a lower phase compared to the period before 2001. During the period, the Gini coefficient is decreased (income distribution improved) and the Pareto coefficient is increased (a greater proportion of income was acquired by the elite). Furthermore, the percentage share of labor who earns under the minimum wage has increased from $32\%$ in 2002 to $45\%$ in 2019. Moreover, when we compare the population share from under $x_t$ to the minimum wage, the ratio increases from $36\%$ in 2002 to almost $52\%$ in 2019. Hence, the decrease in Gini and the increase in Pareto indicates that income is transferred from middle-income groups to lower and upper-income groups. This is another important result deduced from our investigation.

 \subsection{Temporal Variation of The Share of Labour Force and The Employment Rate }

In order to obtain the temporal variation of the share of labor and the employment rate,
we need to calculate the fractions of the share of labor in total output ($u$) and employment rate ($v$) which fits to Gompertz distribution with reference to individual income to obtain the Goodwin cycle. By using unemployment income threshold $x_d<x_t$, the terms $u$ and $v$ can be calculated via
\eq{denk22}{
v=1-\mathcal{F}(x_d),\quad u=\mathcal{F}_1(x_t).
 }

For the aim of determination of the unemployment rate, $x_d$ threshold value has been assumed to be $50\%$ of the annual minimum wage. The $50\%$ rate was determined in the content of the law which unemployment benefit has started to be implemented in the Turkish economy. The amount paid by the state to an unemployed person for survival is calculated as $50\%$ of the average of the last four months' minimum wage. Therefore, it is a reasonable assumption that those whose annual income is $50\%$ or less of the annual minimum wage are considered unemployed, based on their annual income. Thus, those who earn annual income below $x_t$ threshold value of total individuals are included in laborers as low or middle income class, while those who earn less than $50\%$ of the annual minimum wage within this population are defined as effectively unemployed. The effective unemployment is a concept that includes people who have lost their for finding  a job in addition to the unemployed, and those who can work less than they want or need by being employed part-time. 
 
 \begin{figure}[h]
\center
\epsfig{file=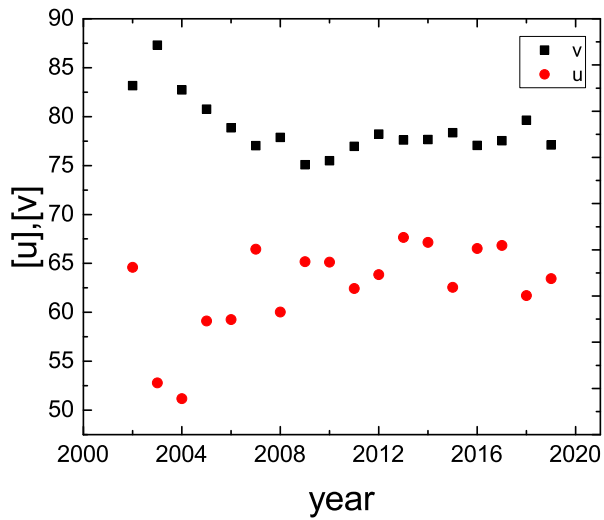, width=8cm}
\caption{The share of labor in total output and employment rate (2002-2019) }\label{sek2}
\end{figure}

In this regard, our calculations from raw data show that, the population rate under $x_d$  which also defines the effective unemployment varies around $22\%$, the population below the minimum wage varies around $37\%$, and the population below $x_t$ which defines the laborers varies around $88\%$. So, the Pareto part of the population that we study varies around $12\%$.

The temporal variation of $u$ and $v$ can be seen in Fig. \ref{sek2}. 
According to Fig. \ref{sek2}, the employment rate decreases steadily from 2003 until 2009, and after 2009 it follows a stable course around $78\%$. In the context of the Turkish economy, there has not been a recovery in employment rates, especially after the crisis. This situation is consistent with the jobless growth phenomenon discussed in the post-2002 period for the Turkish economy. The fact that there has not been a significant improvement in employment rates after 2007, but the country has historically displayed the highest growth rates in this period shows the consistency of the findings of our study with the economic reality. The share of workers in the total product, on the other hand, has been decreased dramatically between 2002 and 2004 and fluctuated around $65\%$ for the other period. The years in which the share of labor force decreases in the Turkish economy (2008, 2011, 2015, 2018) can be observed in the calculations. The global financial crisis of 2008 affected the Turkish economy, as well as the world. However, 2011, 2015 and 2018 are the general election years for the Turkish economy. Therefore, even though there are too many political and external factors during this period, yet there is a specific and regular pattern of accumulation. When we examine the recovery periods of the share of labor force in total output, we can indicate that all recovery periods start with a general election and regress to the previous levels until the next election.

\subsection{Cycles in the Income and Employment Data of Turkey}
 
\begin{figure}[h]
\center
\epsfig{file=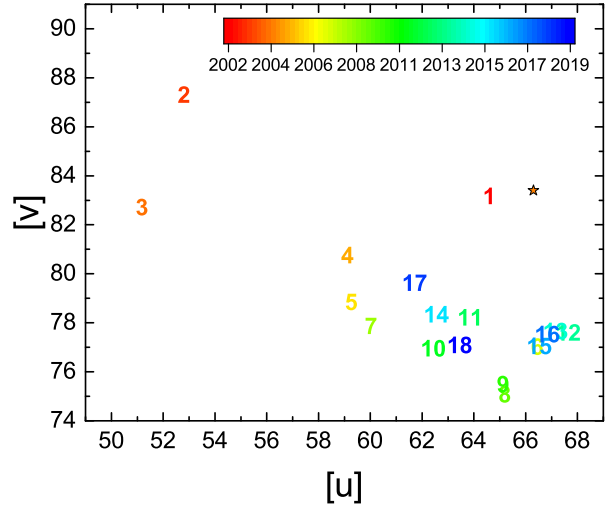, width=8cm}
\caption{The Goodwin cycle for Turkish economy (2002-2019). The star denotes the calculated center of the Goodwin cycle. }\label{sek3}
\end{figure}

The cyclic behavior of income-employment data for Turkish economy between the period $2002-2019$ can be seen in Fig. \ref{sek3}.    

The years between 2002-2019 of the Turkish economy was a period when the single-party government continued uninterruptedly, two major crises occurred on a global and local scale, a social rebellion (Gezi Park) and a coup attempt and many elections (six general elections [2002, 2007, 2001, 2015/June, 2015/November and 2018], 4 mayoral elections [2004, 2009, 2014, 2019] and three referenda [2007, 2010, 2017]). The emergence of cycle or semicircular structures throughout the entire period is an indication of the emergence of an accumulation structure independent of external phenomena.

Fig. \ref{sek3} shows the Goodwin cycle for Turkish economy in 2002-2019 period.  A counter-clockwise cycle can be clearly observed in the Turkish economy for the years 2002-2013, which can be seen in Fig. \ref{sek3}. This short-term cycle brings qualitative empirical support to the Goodwin cycles at least as far as the Turkey data is concerned.

In 2001, Turkish economy had the worst internal crisis ever seen. Thus, the period that we examine here is a recovery period relatively. After the initial point, the year 2002, we can see that there is a strong period of wage suppression, from Fig. \ref{sek2}. The $u$ decreases dramatically until 2004. Although, Gini coefficient is decreased slightly the Pareto coefficient increased and this situation may be an indicator of the fact that income is transferred from middle income to lower income and upper income groups in these years (please see Table \ref{tbl_1}). In the cycle, the year 2007 deviates significantly from the cycle. Hence, 2007 is the beginning of a global crisis period and this breaking year can be observed in the cycles we have calculated in the study. This situation is consistent with the devastating impacts of crisis, or war times on economic activity also results in an increase in inflation, contraction in real wages and purchasing power. High inflation can cause serious changes in the sharing of wages, despite policies to protect the lowest wages more from inflation. For example, in both world wars, the decrease in the share capital received from national income (capital / income ratio) and the decrease in wage inequalities occurred simultaneously \cite{ref63,ref64}. In the post-crisis period, firstly, the employment rate decreased in spite of a recovery in share of labor force. Then, the short-term cycle that we indicate ends in 2013 with the highest share of labor force in total output and relatively low employment rate during this cycle. During the period we analyzed, the inverse relationship between the Gini and Pareto coefficients, excluding the crisis years, shows us that a resource transfer from middle-income to low and high-income earners was applied as a macroeconomic policy in the Turkish economy. This policy choice, obviously increased the income inequalities in Turkey. Additionally, despite a slight improvement in the Gini coefficient after 2013, the Pareto coefficient remains almost constant. 2018 is again a period of internal crisis for the Turkish economy and there is a decreasing Pareto coefficient with increasing Gini. In 2019, relations returned to the beginning of the period, with a decreasing Gini and an increasing Pareto coefficient. 

\begin{figure}[h]
\center
\epsfig{file=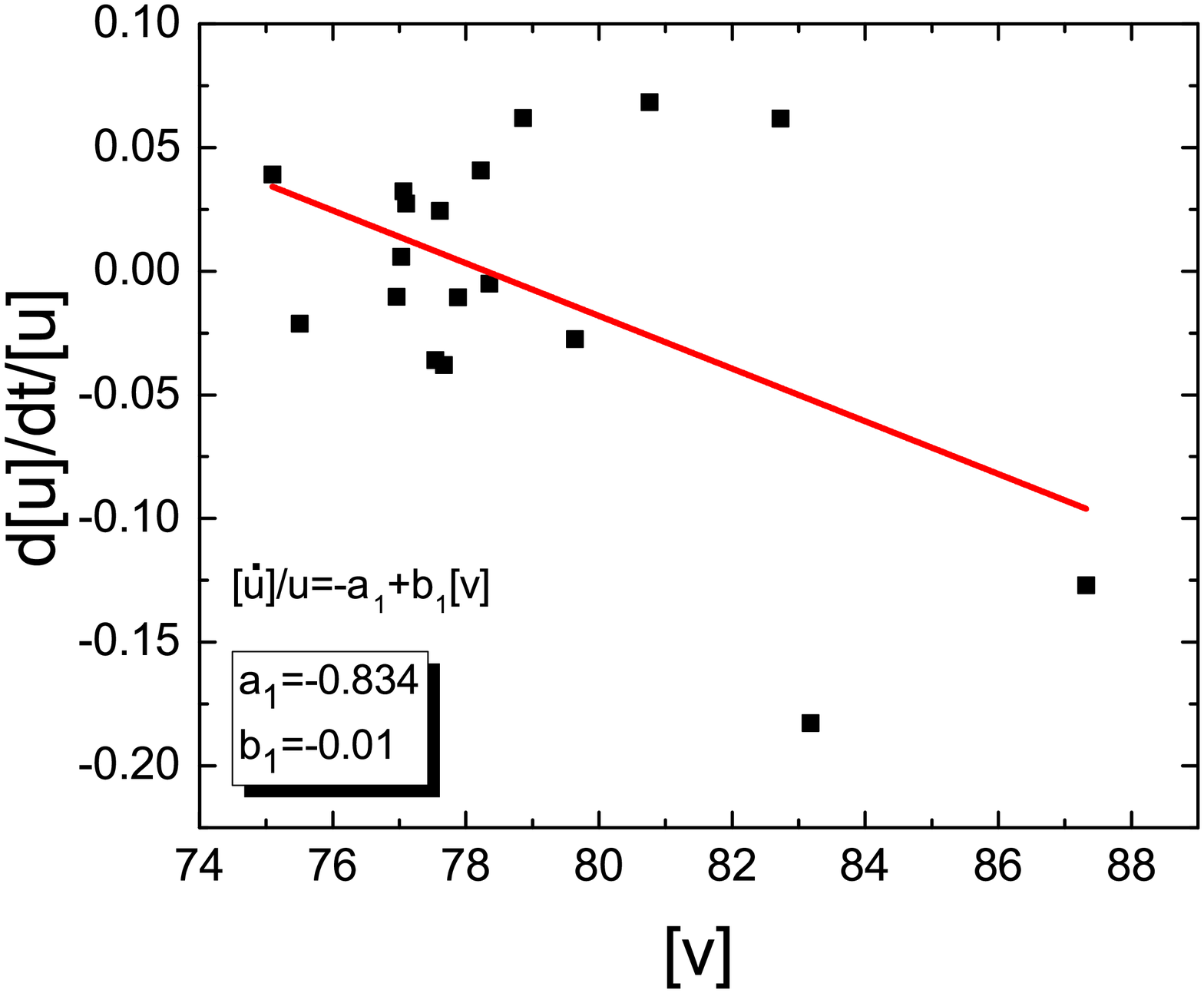, width=6.0cm}
\epsfig{file=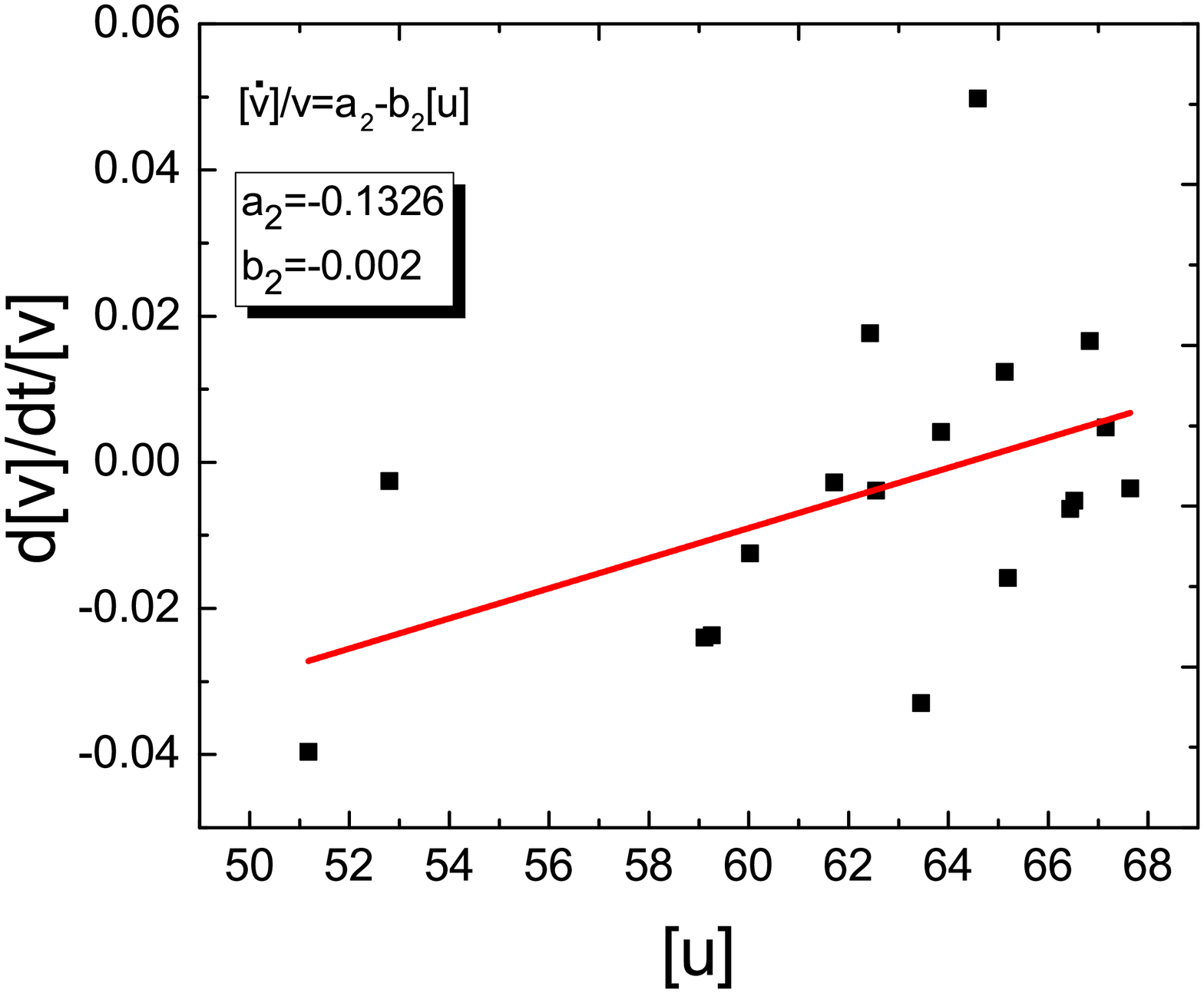, width=6.0cm}
\caption{Fitting of the Goodwin cycle data to the LV  coefficients.}\label{sek4}
\end{figure}

In order to decide whether the calculated coefficients in Eq. \re{denk2}  
produce meaningful Goodwin cycles or not, we fit the  Goodwin cycle data seen in Fig. \ref{sek3} by taking time derivatives of the $u$ and $v$ numerically. Behavior of Eq. \re{denk2} can be seen in Fig. \re{sek4}, as well as fit results of the coefficients in Eq. \re{denk5}.

In Fig. \ref{sek4}, we can clearly identify a general tendency for the observational points to decrease in Fig. \ref{sek4} (a) and to increasing trend  in Fig. \ref{sek4} (b) . The aim of this graphs are to use the fitting procedure to estimate the parameters of the dynamic differential equations. By using these coefficients, we can calculate the center and the period of the cycle from Eqs. \re{denk6} and \re{denk7}.

Behaviors depicted in Fig. \ref{sek4} and results of the fit verify that a  Goodwin cycle occurs. We calculate from Eq. \re{denk6} 
the center of the cycle as $u_c=66.29\%$  for the share of labor force and $v_c=83.40\%$ for the employment rate. Besides we found from Eq. \re{denk7} that the period of the cycle  is $T=18.89$ years.

Note also that, Fig. \ref{sek4} clearly shows the quarters of the Goodwin cycle. The cycle starts at point $v=83.175$ and $u=64.593$, which are calculated values from Eq. \re{denk22} for year 2002.  Downward movement occurs firstly in Fig. \ref{sek4} (a). This corresponds to the rising $v$ and decreasing $u$ quarter of the whole cycle. By reaching maximum $v$ point, movement continues upward in the same curve. This is the decreasing $v$ and decreasing $u$ regime. By reaching the condition $du/dt=0$, another regime starts in the cycle: decreasing $v$ and increasing $u$. After reaching the minimum $v$ value, reverse movement starts and this corresponds to the last quarter of the Goodwin cycle: increasing $v$ and increasing $u$. The same conclusion can be drawn from Fig. \ref{sek4} (b) also.

\section{Conclusion}\label{sec_conclusion}

In this study, the Goodwin cycles, based on LV prey-predator models, was tested for Turkish economy in 2002-2019 period with econophysics approach. We compared the historical facts of Turkish economy with our findings to verify our calculations and results.

Firstly, by examining the distributional structure of individual income of Turkish economy, it has been shown by performing fitting processes that the income has a two-regime distributional structure in the Turkish economy in accordance with the literature. As a result, two-regime distribution which was a fit of Gompertz distribution under a specific threshold average value of $x_t\cong  1.870$, in  normalized income data. In addition, we calculate the Pareto coefficient which has an average value $\alpha\cong  2.065$ for Turkish economy in 2002-2019 period. The Gompertz part of distribution represents the majority of the population ($\sim 88\%$), and the Pareto distribution describes the richer part ($\sim12\%$). This is the first study that explains the distributional structure of individual income with GPD for Turkish economy. Moreover, we interpreted the simultaneous decrease in Gini coefficient and the increase in Pareto coefficient as an indicator of income inequality in Turkey during the 2002-2019 period. This opposite movement of these two indicators might be an indicator of income transfer from the middle-income group to the low and high-income groups.

Secondly, the LV treatment of Goodwin model has been investigated by data regarding the Turkish economy. Obtained average values of $v$ and $u$ are as $v\cong  77.5\%, u\cong  62.5\%$, which are strongly consistent with the post-depression jobless growth of Turkish economy.  There are two orbital centers obtained visually. One is almost a clear cycle for 2002-2013 period which also indicates the 2007/8 breakdown. The other is rather a political and a more closed cycle between 2014 and 2019.  Besides, by performing fitting on the cycle data, we verify that Goodwin cycle is obtained consistently with the LV formulation of Goodwin cycle. We found the center of the cycle as $u_c=66.29\%$ for the share of labor  and  $v_c=83.40\%$  for the employment rate and we found  $T=18.99$ years for the period of the cycle. The contribution of the study is to estimate for the first time the Goodwin model based on micro data by using econophysics approach on the individual data in the HBS for the Turkish economy.

Consequently, the GPD provides an adequate, analytically simple and consistent result for income distribution data from Turkey. The Goodwin cycle that we estimated depending on GPD of individual income in HBS data again provides a coherent and qualitative representation of economic relations in Turkish economy in 2002-2019 period. Consequently, the increase in income inequality in Turkey can also be observed with Goodwin cycle that we have estimated during this period.

We hope that the results
obtained in this work may be beneficial form econophysical point of view.


\end{document}